\documentclass[twocolumn,showpacs,preprintnumbers,nofootinbib,amsmath,amssymb]{revtex4}

\usepackage{graphicx}
\usepackage{dcolumn}
\usepackage{bm}
\usepackage{epsfig}

\begin{document}

\preprint{DESY~15-104\hspace{13.5cm} ISSN~0418-9833}

\title{\boldmath Complete Nonrelativistic-QCD Prediction for Prompt Double
$J/\psi$ Hadroproduction}

\author{Zhi-Guo He}
\author{Bernd A. Kniehl}
\affiliation{{II.} Institut f\"ur Theoretische Physik, Universit\"at Hamburg,
Luruper Chaussee 149, 22761 Hamburg, Germany}

\date{\today}

\begin{abstract}
We perform a complete study of prompt double $J/\psi$ hadroproduction at
leading order in the nonrelativistic-QCD factorization framework by including
all possible pairings of the $c\bar{c}$ Fock states ${}^1\!S_0^{[8]}$,
${}^3\!S_1^{[1,8]}$, and ${}^3\!P_J^{[1,8]}$ with $J=0,1,2$.
We find that the ${}^1\!S_0^{[8]}$ and ${}^3\!P_J^{[8]}$ channels of
$J/\psi$ and $\psi^\prime$ production and the ${}^3\!P_J^{[1]}$ and
${}^3\!S_1^{[8]}$ channels of $\chi_{cJ}$ production, which have been overlooked
so far, greatly dominate at large invariant masses and rapidity separations of
the $J/\psi$ pair, and that their inclusion nearly fills the large gap
between previous incomplete predictions within the color-singlet model and the
recent measurement by the CMS Collaboration at the CERN LHC, leaving room for
next-to-leading-order corrections of typical size.
\end{abstract}

\maketitle

The nonrelativistic QCD (NRQCD) \cite{Caswell:1985ui} factorization formalism,
introduced two decades ago in a seminal work by Bodwin, Braaten, and Lepage
\cite{Bodwin:1994jh}, nowadays is the only game in town for the
theoretical description of heavy-quarkonium production and decay, and its
experimental verification is generally considered to be among the most urgent
tasks of heavy-quarkonium physics \cite{Brambilla:2010cs}.
The production cross sections and decay rates are separated into
process-dependent short-distance coefficients (SDCs), calculated by expansion
in the strong-coupling constant $\alpha_s$, and universal long-distance matrix
elements (LDMEs), which are strongly ordered in size by velocity ($v$) scaling
rules \cite{Lepage:1992tx}.
The heavy-quark pair may appear in any Fock state $n={}^{2S+1}\!L_{J}^{[a]}$,
both as color singlet (CS) $a=1$ and color octet (CO) $a=8$, thus giving rise
to the CO mechanism (COM), while, in the traditional CS model, it is restricted
to the CS state sharing the spectroscopic quantum numbers ${}^{2S+1}\!L_J$
with the physical quarkonium state considered.
Despite its aesthetic simplicity and theoretical rigor, consolidated very
recently by an all-order proof \cite{Nayak:2015qca}, NRQCD factorization has
reached the crossroads because the predicted universality of the LDMEs is
challenged \cite{Butenschoen:2012qr} by recent measurements of $J/\psi$
polarization \cite{Butenschoen:2012px} and $\eta_c$ yield
\cite{Butenschoen:2014dra}, which is in the very focus at the CERN LHC.

\begin{figure}
\centering
\begin{tabular}{c}
\includegraphics[scale=0.40]{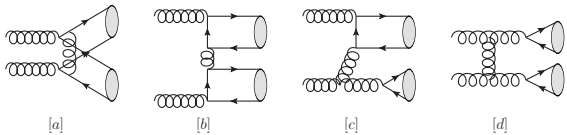}
\end{tabular}
\caption{Typical Feynman diagrams for $gg\to c\bar{c}(m)c\bar{c}(n)$:
(a) nonfragmentation type~I,
(b) nonfragmentation type~II,
(c) single-fragmentation-like,
(d) double-fragmentation-like.}
\label{Feynman}
\end{figure}

Our Letter addresses another burning problem of NRQCD, namely, its seeming
failure to describe recent measurements of prompt double $J/\psi$
hadroproduction performed by the LHCb \cite{Aaij:2011yc} and
CMS \cite{Khachatryan:2014iia} Collaborations at the LHC, and the 
D0 Collaboration \cite{Abazov:2014qba} at the Fermilab Tevatron. 
This is a particularly sensitive testing ground for NRQCD factorization, which
takes effect there twice, and a topic of old vintage, pioneered by
Ref.~\cite{Barger:1995vx} in 1995, which has attracted considerable
theoretical interest since then (see, e.g.,
Refs.~\cite{Qiao:2002rh,Li:2009ug,Qiao:2009kg,Ko:2010xy,Berezhnoy:2011xy,%
Li:2013csa,Sun:2014gca}), but is much less advanced than single $J/\psi$
production.
So far, only the CS contribution due to $gg\to2c\bar{c}({}^3\!S_1^{[1]})$ and
the CO contribution due to $gg\to2c\bar{c}({}^3\!S_1^{[8]})$, which resembles
double fragmentation [see Fig.~\ref{Feynman}(d)], have been studied for
direct $J/\psi$ production and also for the feed down from
$\psi^\prime$ mesons, which requires no extra calculation
\cite{Barger:1995vx,Qiao:2002rh,Li:2009ug,Qiao:2009kg,Ko:2010xy,%
Berezhnoy:2011xy,Li:2013csa,Sun:2014gca}.
These calculations of prompt double $J/\psi$ production, which we henceforth
denote as CS${}^\ast$ and CO${}^\ast$, respectively, are incomplete because
they lack the ${}^1\!S_0^{[8]}$ and ${}^3\!P_J^{[8]}$ contributions to $J/\psi$
and $\psi^\prime$ production and the ${}^3\!P_J^{[1]}$ and ${}^3\!S_1^{[8]}$
contributions to $\chi_{cJ}$ production, where $J=0,1,2$.
Interestingly, $J/\psi+\chi_{cJ}$ production is forbidden at
$\mathcal{O}(\alpha_s^4)$ in the CS model by $CP$ conservation, while it is
enabled by the COM of NRQCD.
Thus, we are led to include a total of $\binom{8}{2}-3=25$ different pairings
of $c\bar{c}$ Fock states altogether, as indicated in Table~\ref{p_T},
out of which only 2 have been considered so far.
In our Letter, we demonstrate that NRQCD factorization may be reconciled with
the experimental data \cite{Aaij:2011yc,Khachatryan:2014iia,Abazov:2014qba},
leaving room for typical next-to-leading-order (NLO) corrections, if the
previously neglected CO and feed-down channels are properly included.
We thus add another crucial piece of information to the tantalizing tale of
NRQCD factorization \cite{Bodwin:1994jh} and point into a new direction, namely
the relative $\mathcal{O}(\alpha_s)$ corrections to the next-to-leading-power
(NLP) and next-to-next-to-leading-power (NNLP) CO processes of prompt double
$J/\psi$ hadroproduction to be identified below.
If their inclusion turned out to bring the NRQCD prediction in agreement with
the LHC data, which we deem very likely for reasons explained below, this would
be an important milestone in the verification of the COM, which is a key
prediction of NRQCD factorization.
Owing to the predicted LDME universality, double $J/\psi$ production will then
also yield independent constraints on yield and polarization of single
$J/\psi$ production.

Our Letter also suggests a solution to another important QCD problem of general
interest \cite{Kom:2011bd}, namely the double-parton-scattering (DPS) surplus
observed by the D0 Collaboration \cite{Abazov:2014qba}.
In fact, their result for
$\sigma_\mathrm{eff}=(\sigma_{J/\psi})^2/\sigma_\mathrm{DPS}$ is considerably
smaller than the findings by other experiments \cite{Abazov:2014qba}.
The increase of the single-parton-scattering (SPS) portion
$\sigma_\mathrm{SPS}$ due to our completion of the NRQCD prediction results in
a reduction of $\sigma_\mathrm{DPS}$, which in turn increases
$\sigma_\mathrm{eff}$ and so places it in the ball park of other
determinations.

So far, the experimental data
\cite{Aaij:2011yc,Khachatryan:2014iia,Abazov:2014qba},
which come as total cross sections $\sigma_\mathrm{tot}$ and distributions in
the invariant mass $M$, the transverse momentum $P_T$, and the rapidity ($y$)
separation $|\Delta y|$ of the $J/\psi$ pair, have mostly been compared with
CS${}^\ast$ predictions, which dominate for small values of the $J/\psi$
transverse momentum $p_T$ \cite{Qiao:2002rh,Li:2009ug,Qiao:2009kg},
while the CO${}^\ast$ contributions take over in the large-$p_T$ region, for
$p_T\agt16$~GeV at the LHC \cite{Ko:2010xy}.
In the LHCb \cite{Aaij:2011yc} case, the CS${}^\ast$ prediction for
$\sigma_\mathrm{tot}$, which receives a moderate enhancement of relative order
$\mathcal{O}(\alpha_s)$ of about 10\% \cite{Sun:2014gca}, is compatible with
the measurement, but the one for the distribution $d\sigma/dM$ significantly
overshoots the data points close to the $J/\psi$ pair production threshold,
even after including the negative corrections of relative order
$\mathcal{O}(v^2)$, which are about $-23\%$ \cite{Li:2013csa}.
In the CMS \cite{Khachatryan:2014iia} case, the CS${}^\ast$ prediction for
$\sigma_\mathrm{tot}$, which is enhanced by more than 1 order of magnitude by
relative $\mathcal{O}(\alpha_s)$ corrections \cite{Sun:2014gca}, can only
account for about $2/3$ of the measurement, the one for the distribution
$d\sigma/dP_T$ significantly differs from the measurement as for the line shape,
and the one for $d\sigma/dM$ dramatically undershoots the measurement, by 4
orders of magnitude in the large-$M$ region, for $M>35$~GeV.
This enormous discrepancy seriously jeopardizes the validity of NRQCD
factorization \cite{Bodwin:1994jh}, and it is an important task of general
interest to perform a systematic study of all the contributing channels, which
is the very purpose of our Letter.
In the D0 \cite{Abazov:2014qba} case, there is also a large gap between the
CS${}^\ast$ prediction and the experimental result for the SPS cross section
\cite{Abazov:2014qba}.

Owing to the factorization theorems of the QCD parton model and NRQCD, the
prompt double $J/\psi$ hadroproduction cross section may be evaluated as
\begin{eqnarray}\label{xs}
\lefteqn{d\sigma(AB\to 2J/\psi+X) =\sum_{i,j,m,n,H_1,H_2}\int dx_1 d x_2}
\nonumber\\
&&{}\times f_{i/A}(x_1)f_{j/B}(x_2)d\hat{\sigma}(ij\to c\bar{c}(m)c\bar{c}(n)+X)
\nonumber\\
&&{}\times \langle\mathcal{O}^{H_1}(m)\rangle\mathrm{Br}(H_1\to J/\psi+X)
\nonumber\\
&&{}\times\langle\mathcal{O}^{H_2}(n)\rangle\mathrm{Br}(H_2\to J/\psi+X),
\end{eqnarray}
where $f_{i/A}(x)$ is the parton distribution function (PDF) of parton $i$ in
hadron $A$,
$d\hat{\sigma}[ij\to c\bar{c}(m)c\bar{c}(n)+X]$ is the SDC,
$\langle\mathcal{O}^H(m)\rangle$ is the LDME of
$H=J/\psi,\chi_{cJ},\psi^\prime$, and $\mathrm{Br}(H\to J/\psi+X)$ is the
branching fraction with the understanding that
$\mathrm{Br}(H\to J/\psi+X)=1$ if $H=J/\psi$.
Since the $q\bar{q}$-initiated subprocesses are greatly suppressed by the
light-quark PDFs \cite{Qiao:2009kg}, we concentrate on $gg$ fusion.
Because of the smallness of $\mathrm{Br}(\chi_{c0}\to J/\psi\gamma)=1.27\%$
\cite{Agashe:2014kda}, we neglect the contributions from $H=\chi_{c0}$.
Our analytic results for the CS${}^\ast$ and CO${}^\ast$ channels agree with
the literature \cite{Qiao:2002rh,Ko:2010xy}.

There is a total of 72 Feynman diagrams contributing to the generic partonic
subprocess $gg\to c\bar{c}(m)c\bar{c}(n)$, and representative ones are depicted
in Fig.~\ref{Feynman}.
For given $m$ and $n$, not all of them contribute due to $J^{PC}$ conservation.
According to the scaling $d\sigma/dp_T^2\propto1/p_T^N$ and the topologies of
the contributing Feynman diagrams [see Figs.~\ref{Feynman}(a)--(d)], we divide
the partonic subprocesses into 4 categories:
(i) NNLP-I, with $N=8$, including
$m={}^3\!S_1^{[1]}$ and $n={}^3\!S_1^{[1,8]},{}^1\!S_0^{[8]},{}^3\!P_J^{[8]}$;
(ii) NNLP-II, with $N=8$, too, including
$m,n={}^1\!S_0^{[8]},{}^3\!P_J^{[8]},{}^3\!P_J^{[1]}$;
(iii) NLP, with $N=6$, including $m={}^3\!S_1^{[8]}$ and
$n={}^1\!S_0^{[8]},{}^3\!P_J^{[8]},{}^3\!P_J^{[1]}$; and
(iv) leading power (LP), with $N=4$, including $m=n={}^3\!S_1^{[8]}$.
While the NNLP-I and NNPL-II subprocesses exhibit the same $p_T$ scaling, they
differ by the topologies of the respective Feynman diagrams.
In the latter case, these are the diffractionlike ones as in
Fig.~\ref{Feynman}(b), which allow for large values of $|\Delta y|$ and thus
for an enhancement of the cross section at large values of $M$. 
Also taking into account the scaling with $v$ of the LDMEs and noticing that
$\mathrm{Br}(\chi_{c1,2}\to J/\psi\gamma)=\mathcal{O}(v^2)$ numerically, we
roughly estimate the relative importance of each channel at large values of
$p_T$ as summarized in Table~\ref{p_T}.

\begin{table}
\caption{Scaling with $p_T$ and $v$ of $d\sigma/dp_T^2$ for
$gg\to c\bar{c}(m)c\bar{c}(n)$ times the respective LDMEs and branching
fractions for the relevant pairings $(m,n)$ of $c\bar{c}$ Fock states.
Note that ${}^3\!P_J^{[1]}$ are counted separately for $J=0,1,2$.}
\begin{tabular}{|c|c|c|c|c|c|}
\hline
$(m,n)$ & ${}^3\!S_1^{[1]}$ & ${}^3\!S_1^{[8]}$ & ${}^1\!S_0^{[8]}$ &
${}^3\!P_J^{[8]}$ & ${}^3\!P_J^{[1]}$ \\
\hline
${}^3\!S_1^{[1]}$ & $1/p_T^8$ & $v^4/p_T^8$ & $v^3/p_T^8$ & $v^4/p_T^8$ & 0 \\
\hline
${}^3\!S_1^{[8]}$ & $\cdots$ & $v^8/p_T^4$ & $v^7/p_T^6$ & $v^8/p_T^6$ & $v^8/p_T^6$ \\
\hline
${}^1\!S_0^{[8]}$ & $\cdots$ & $\cdots$ & $v^6/p_T^8$ & $v^7/p_T^8$ & $v^7/p_T^8$ \\
\hline
${}^3\!P_J^{[8]}$ & $\cdots$ & $\cdots$ & $\cdots$ & $v^8/p_T^8$ & $v^8/p_T^8$ \\
\hline
${}^3\!P_J^{[1]}$ & $\cdots$ & $\cdots$ & $\cdots$ & $\cdots$ & $v^8/p_T^8$ \\
\hline
\end{tabular}\label{p_T}
\end{table}


We work at leading order (LO) in the fixed-flavor-number scheme with 3
massless quark flavors and a charm-quark mass of $m_c=1.5$~GeV.
We use the LO formula for $\alpha_s^{(4)}(\mu_r)$ with asymptotic scale
parameter $\Lambda^{(4)}=192$~MeV \cite{Lai:1999wy} and the CTEQ5L set of LO
proton PDFs \cite{Lai:1999wy}.
We choose the renormalization and factorization scales as
$\mu_r=\mu_f=\xi\sqrt{(4m_c)^2+p_T^2}$ and vary $\xi$ between $1/2$ and 2
about the default value 1 to estimate the theoretical uncertainty.
As for the LDMEs of the $J/\psi$, $\chi_{cJ}$, and $\psi^\prime$ mesons, we adopt
the CS values from Ref.~\cite{Eichten:1995ch}, evaluated using the
Buchm\"uller-Tye potential, and the CO values from Ref.~\cite{Braaten:1999qk},
fitted to single $J/\psi$ hadroproduction data at LO in NRQCD.
Because of the strong correlations between
$\langle\mathcal{O}^H({}^1\!S_0^{[8]})\rangle$ and
$\langle\mathcal{O}^H({}^3\!P_0^{[8]})\rangle$ for $H=J/\psi,\psi^\prime$, only
the linear combinations
$M_r^H=\langle\mathcal{O}^H({}^1\!S_0^{[8]})\rangle
+r\langle\mathcal{O}^H({}^3\!P_0^{[8]})\rangle/m_c^2$ could
be determined in Ref.~\cite{Braaten:1999qk}.
Fortunately, these correlations are very similar in prompt double $J/\psi$
hadroproduction via the NNLP-II and NLP subprocesses.
We use $\mathrm{Br}(\chi_{c1}\to J/\psi\gamma)=33.9\%$,
$\mathrm{Br}(\chi_{c2}\to J/\psi\gamma)=19.2\%$, and
$\mathrm{Br}(\psi^\prime\to J/\psi+X)=60.9\%$ \cite{Agashe:2014kda}.

Prior to performing detailed comparisons with measurements, we expose some
general features of our results.
(a) Among the NNLP-I subprocesses, no kinematic enhancements are found relative
to the CS${}^\ast$ channel, so that all the other channels are suppressed as
$\mathcal{O}(v^3)$ by the LDMEs.
(b) Although the $p_T$ scaling of the NNLP-II subprocesses is as unfavorable as
that of the NNLP-I ones, their SDCs may be about 50--200 times larger than that
of the CS${}^\ast$ channel.
(c) The contribution of the NLP subprocesses may also exceed that of the
CS${}^\ast$ channel, e.g., for $p_T>20$~GeV under CMS kinematic conditions.
(d) At large values of $M$, the $M$ scalings and the corresponding $p_T$
scalings of the 4 types of subprocesses are the same, but the differential
cross sections $d\sigma/dM$ of the NNLP-II, NLP, and LP subprocesses may be
more than 1 order of magnitude larger than that of the CS${}^\ast$ channel.
Observations (b)--(d) indicate that the combination of the CS${}^\ast$ and
CO${}^\ast$ contributions, $d\sigma^{\ast}$, may not be a good approximation to
the full NRQCD result, $d\sigma$, especially at large values of $M$.
(e) As expected from identical-boson symmetry and the $J/\psi+\chi_{cJ}$
suppression mentioned above, the relative importance of the $\chi_{cJ}$
($\psi^\prime$) feed-down contribution is reduced (increased) with respect to
prompt single $J/\psi$ hadroproduction.

\begin{figure}
\centering
\begin{tabular}{c}
\includegraphics[scale=0.80]{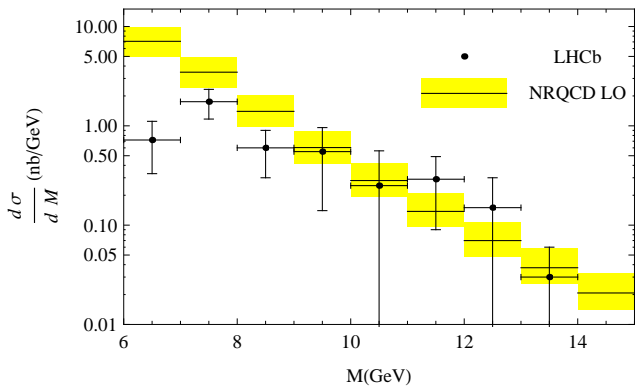}
\end{tabular}
\caption{The $M$ distribution of prompt double $J/\psi$ hadroproduction
measured by LHCb \cite{Aaij:2011yc}
is compared to the full LO NRQCD prediction (solid lines).
The theoretical uncertainty is indicated by the shaded (yellow) bands.}
\label{lhcmjj}
\end{figure}

The LHCb Collaboration \cite{Aaij:2011yc} measured $\sigma_\mathrm{tot}$ at
center-of-mass (CM) energy $\sqrt{s}=7$~TeV requiring $p_T<10$~GeV and
$2.0<y<4.5$ for each of the $J/\psi$ mesons to find
$\sigma_\mathrm{tot}^\mathrm{LHCb}=(5.1\pm1.0\pm1.1)$~nb.
Our corresponding LO NRQCD predictions are
$\sigma_\mathrm{tot}^{\ast}=12.2^{+4.8}_{-3.8}$~nb,
which is somewhat larger than in Refs.~\cite{Li:2013csa,Sun:2014gca} because of
different choices of $m_c$, LDMEs, PDFs, and scales, and
$\sigma_\mathrm{tot}=13.2^{+5.2}_{-4.1}$~nb, which is about 2.6 times larger
than the LHCb result.
To better understand the origin of this excess, we consider in
Fig.~\ref{lhcmjj} the LHCb and full LO NRQCD results differential in $M$.
We observe that the theoretical prediction systematically overshoots the
experimental data in the threshold region, where $M\alt9$~GeV, while there is
nice agreement for larger values of $M$.
Near the $J/\psi$ pair production threshold, multiple soft-gluon emissions
spoil the perturbative treatment, relativistic corrections are nonnegligible
\cite{Martynenko:2012tf}, and $\sigma_\mathrm{tot}\propto m_c^{-8}$
\cite{Li:2013csa}, which amplifies the theoretical uncertainty.
All these effects are likely to render a LO NRQCD analysis inappropriate there.

The CMS data \cite{Khachatryan:2014iia} were taken at the same CM energy, but
are subject to a $y$-dependent low-$p_T$ cut and cover a more central $y$ range
than the LHCb data,
as specified in Eq.~(3.3) of Ref.~\cite{Khachatryan:2014iia}.
They yield
$\sigma_\mathrm{tot}^\mathrm{CMS}=(1.49\pm0.07\pm0.13)$~nb.
Our LO NRQCD predictions are
$\sigma_\mathrm{tot}^{\ast}=0.10^{+0.05}_{-0.03}$~nb and
$\sigma_\mathrm{tot}=0.15^{+0.08}_{-0.05}$~nb, which is still 1 order of
magnitude smaller than the CMS measurement.
The NNLP-I, NNLP-II, NLP, and LP contributions to the central value of
$\sigma_\mathrm{tot}$ are 97, 13, 27, and 14~fb, respectively.
I.e., over 36\% of $\sigma_\mathrm{tot}$ is made up
by the NNLP-II, NLP, and LP processes;
about one half of this contribution comes as feed down from $\chi_{cJ}$ mesons,
via $J/\psi+\chi_{cJ}$ and $\chi_{cJ}+\chi_{cJ}$.
Therefore, the CS${}^\ast$ approximation is bound to be
insufficient, even after including the $\mathcal{O}(\alpha_s)$ corrections
\cite{Sun:2014gca}.
To substantiate this statement, we also consider the scaling
$d\sigma/dp_T^2\propto1/p_T^N$.
In the CS${}^\ast$ channel, we have $N=8$ at LO and $N=6$ at NLO
\cite{Sun:2014gca,Lansberg:2013qka}.
Similarly, the NNLP-II and NLP processes at NLO are expected to have $N=6$ and
$N=4$, respectively, and are thus likely to produce sizable enhancements as
well.
Correction factors of 5--10, which appear plausible, would eliminate the
discrepancy between the CMS measurement of $\sigma_\mathrm{tot}$ and the
NRQCD prediction.

\begin{figure}
\centering
\begin{tabular}{c}
\includegraphics[scale=0.80]{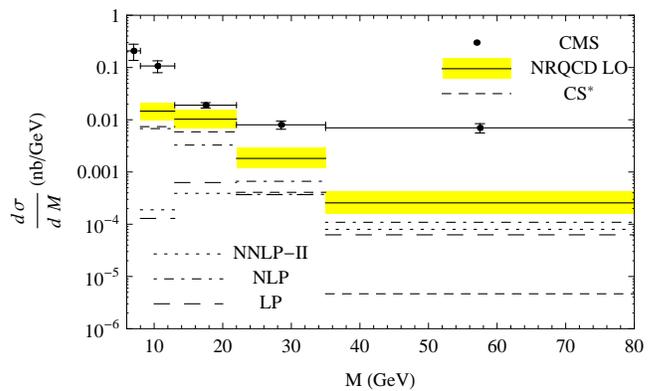}
\end{tabular}
\caption{The $M$ distribution of prompt double $J/\psi$ hadroproduction
measured by CMS \cite{Khachatryan:2014iia}
is compared to the full LO NRQCD
prediction (solid lines), its NNLP-II (dotted lines), NLP (dot-dashed lines),
and LP (long-dashed lines) components, and the LO CS${}^\ast$ contribution
(dashed lines).
The theoretical uncertainty in the LO NRQCD prediction is indicated by the
shaded (yellow) bands.}
\label{cmsmjj}
\end{figure}

The CMS Collaboration also measured the differential cross section in bins of
$M$ and $|\Delta y|$.
As mentioned above, the $\mathcal{O}(\alpha_s)$-corrected CS${}^\ast$ prediction
for $d\sigma/dM$ \cite{Sun:2014gca} dramatically undershoots the CMS data at
large values of $M$, by about 2 and 4 orders of magnitude in the two outmost
bins $22~\mathrm{GeV}<M<35~\mathrm{GeV}$ and
$35~\mathrm{GeV}<M<80~\mathrm{GeV}$, respectively.
In Fig.~\ref{cmsmjj}, we confront these CMS data with our full LO NRQCD result
also showing the LO CS${}^\ast$, NNLP-II, NLP, and LP contributions for
reference.
We observe that the previously neglected NRQCD contributions greatly help to
fill the gap between data and theory.
After their inclusion, the LO NRQCD predictions are only about 4 and 30 times
smaller than the CMS data in the last two bins, where the NNLP-II, NLP, and LP
processes are approximately equally important.

\begin{figure}
\centering
\begin{tabular}{c}
\includegraphics[scale=0.80]{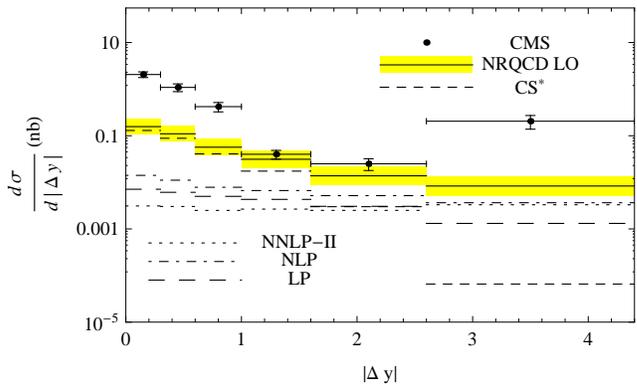}
\end{tabular}
\caption{As in Fig.~\ref{cmsmjj}, but for the $|\Delta y|$ distribution.}
\label{cmsyy}
\end{figure}

At LO, $M$, $p_T$, and $|\Delta y|$ are not independent of each other, but
related by $M=2\sqrt{4m_c^2+p_T^2}\cosh(|\Delta y|/2)$.
Thus, the significant enhancement in the $M$ distribution may be understood
from the $|\Delta y|$ distribution, which is shown in Fig.~\ref{cmsyy}.
We observe from Fig.~\ref{cmsyy} that the CS${}^\ast$ contribution to
$d\sigma/d|\Delta y|$ peaks near $|\Delta y|=0$, which implies that the bulk of
the CS${}^\ast$ contribution to $d\sigma/dM$ at $M\gg2m_{J/\psi}$ arises from
the large-$p_T$ region, with $p_T\approx M/2$, where the cross section is
already very small.
On the other hand, Fig.~\ref{cmsyy} tells us that the inclusion of the residual
LO NRQCD contributions renders the $|\Delta y|$ distribution significantly
broader, which in turn allows for the moderate-$p_T$ region to feed into the
large-$M$ bins so as to increase $d\sigma/dM$ there by orders of magnitude. 
Detailed inspection of the SDCs reveals that the broadening of the
$d\sigma/d|\Delta y|$ peak about $|\Delta y|=0$ is produced by the
pseudodiffractive topologies of Feynman diagrams, with a $t$-channel gluon
exchange, like those in Figs.~\ref{Feynman}(b)--(d).
Although the agreement between the CMS measurement of $d\sigma/dM$ and the
NRQCD prediction is dramatically improved by the inclusion of the missing LO
contributions, there remain appreciable gaps, of roughly 1 order of
magnitude, in the outmost bins in Fig.~\ref{cmsmjj}.
Because of their slower falloff with $p_T$ in connection with the minimum-$p_T$
cut,
the NLO corrections to those new CO and feed-down
contributions, which lie beyond the scope of our present analysis, are likely
to further improve the situation.
That the CMS kinematic conditions give rise to large NLO corrections may also
be understood from the $P_T$ distribution in Fig.~2(c) and Table 4 of
Ref.~\cite{Khachatryan:2014iia} by observing that only 19\% of
$\sigma_\mathrm{tot}$ arise from the lowest bin $P_T<5$~GeV, which includes the
back-to-back situation of the LO calculation, for which $P_T=0$. 
A good part of this bin and all the other bins require the radiation of an
additional parton, which only comes at NLO.
This also explains why the CS${}^\ast$ prediction for $\sigma_\mathrm{tot}$
receives such a sizable $\mathcal{O}(\alpha_s)$ correction \cite{Sun:2014gca}.

The LHCb \cite{Aaij:2011yc} and CMS \cite{Khachatryan:2014iia} measurements
involve both SPS and DPS contributions.
The D0 Collaboration \cite{Abazov:2014qba} attempted to separate them in their
measurement of prompt double $J/\psi$ production in $p\bar{p}$ collisions at
$\sqrt{s}=1.96$~TeV with $p_T>4$~GeV and $|\eta|<2.0$, where $\eta$ is the
$J/\psi$ pseudorapidity, to find
$\sigma_\mathrm{SPS}^\mathrm{D0}=(70\pm6\pm22)$~fb and
$\sigma_\mathrm{DPS}^\mathrm{D0}=(59\pm6\pm22)$~fb.
The central SPS result
exceeds the LO CS${}^\ast$
prediction $\sigma_\mathrm{tot}^{\ast}=51.9$~fb \cite{Qiao:2012wc} by 35\%.
We estimate the residual LO NRQCD contributions, due to the ${}^1\!S_0^{[8]}$
and ${}^3\!P_J^{[8]}$ channels and the feed down from $\chi_{cJ}$ mesons
considered here, to yield a 28\% enhancement, which establishes nice agreement.
The situation might change again after including NLO corrections.
The cutoff-regularized real radiative corrections of relative order
$\mathcal{O}(\alpha_s)$ to the CS${}^\ast$ contribution were considered in
Ref.~\cite{Lansberg:2013qka}.

From the comparisons in three different experimental environments, we conclude
that, in the small-$p_T$ region and away from the $J/\psi$ pair production
threshold, the CS${}^\ast$ calculation provides a reasonable approximation to
the full NRQCD result and acceptable descriptions of the measurements
\cite{Aaij:2011yc,Khachatryan:2014iia,Abazov:2014qba}.
However, at large values of $M$ and $|\Delta y|$, the CS${}^\ast$ contribution
to the full NRQCD prediction is small against those due to the NNLP-II, NLP,
and LP processes, which have been neglected so far.
In fact, their inclusion reduces the gap between the CS${}^\ast$ result and the
CMS data \cite{Khachatryan:2014iia} in the outmost $M$ and $|\Delta y|$ bins by
several orders of magnitude, but leave room for NLO corrections of typical
size.
Should the NLO NRQCD prediction, which is yet to be calculated, agree with the
CMS data, then this would provide strong evidence in favor of the COM.

Prompt double $J/\psi$ hadroproduction also serves as a useful laboratory to
probe the DPS mechanism \cite{Kom:2011bd}.
Reportedly, $(46\pm22)\%$ of the D0 result is due to DPS
\cite{Abazov:2014qba}.
If the determination of the SPS contribution is only based on the CS${}^\ast$
approximation, then the DPS contribution dominates for $|\Delta y|>2.0$ because
of its considerably broader $|\Delta y|$ distribution
\cite{Kom:2011bd}.
However, including the residual NRQCD contributions, due to the NNLP-II, NLP,
and LP processes, on top of the CS${}^\ast$ contribution renders the
$|\Delta y|$ distribution of SPS much broader, as may be seen in
Fig.~\ref{cmsyy} for CMS kinematic conditions, leaving less room for DPS in
agreement with other measurements \cite{Abazov:2014qba}.
In other words, the relative importance of SPS and DPS extracted from
experimental data delicately depends on the quality of the NRQCD
prediction, and any conclusions concerning the significance of DPS are
premature before the NLO corrections to all the relevant channels are taken
into account.

\end{document}